\documentclass[pra,twocolumn]{revtex4-1}

\usepackage[utf8]{inputenc}

\usepackage{amsmath}
\usepackage{amssymb}
\usepackage{mathrsfs}
\usepackage{mathtools}
\usepackage{graphicx}
\usepackage{tikz}
\usepackage{bm}
\usepackage{xcolor}
\usepackage{blindtext}
\usepackage{float}
\usepackage{verbatim}
\usepackage{hyperref}
\usepackage{dsfont}

\newcommand{\be}{\begin{equation}}
\newcommand{\bea}{\begin{eqnarray}}
\newcommand{\ee}{\end{equation}}
\newcommand{\eea}{\end{eqnarray}}
\newcommand{\ben}{\begin{equation*}}
\newcommand{\bean}{\begin{eqnarray*}}
\newcommand{\een}{\end{equation*}}
\newcommand{\eean}{\end{eqnarray*}}
\newcommand{\ba}{\begin{align}}
\newcommand{\ea}{\end{align}}
\newcommand{\ban}{\begin{align*}}
\newcommand{\ean}{\end{align*}}

\newcommand{\ket}[1]{\left|#1\right\rangle}

\graphicspath{{./}{./FiguresHighRes/}}

\begin{document}

\title{Holographically-controlled random numbers from entangled twisted photons}

\author{Michael de Oliveira}
\author{Nicholas Bornman}
\author{Andrew Forbes}

\affiliation{School of Physics, University of the Witwatersrand, Private Bag 3, Wits 2050, South Africa}


\begin{abstract}
\noindent We present a quantum random number generator (QRNG) based on the random outcomes inherent in projective measurements on a superposition of quantum states of light. Firstly, we use multiplexed holograms encoded on a spatial light modulator to spatially map down-converted photons onto a superposition of optical paths. This gives us full digital control of the mapping process which we can tailor to achieve any desired probability distribution. More importantly, we use this method to account for any bias present within our transmission and detection system, forgoing the need for time-consuming and inefficient unbiasing algorithms. Our QRNG achieved a min-entropy of $\text{H}_{\text{min}}=0.9991\pm0.0003$ bits per photon and passed the NIST statistical test suite. Furthermore, we extend our approach to realise a QRNG based on photons entangled in their orbital angular momentum (OAM) degree of freedom. This combination of digital holograms and projective measurements on arbitrary OAM combinations allowed us to generate random numbers with arbitrary distributions, in effect tailoring the system's entropy while maintaining the inherent quantum irreproducibility. Such techniques allow access to the higher-dimensional OAM Hilbert space, opening up an avenue for generating multiple random bits per photon.
\end{abstract}

\maketitle

\section{Introduction}
\label{sec:introduction}

Our human intuition regarding randomness is often faulty. For example, one would think that if one were to draw enough elements from a truly random source, no patterns in the data would emerge. According to Ramsey theory \cite{graham2015rudiments}, however, this is not the case: ideal randomness in huge data sets, in which no conceivable patterns in the data exist, is impossible. This, for example, is why ancient cultures observed the various constellations in the night sky. It is still important for us to understand and use random numbers, in order to make progress in areas such as cryptography~\cite{ekert1991quantum, bennett1992quantum,gisin2002quantum}, numerical simulations including Monte-Carlo integration \cite{metropolis1949monte, rubinstein2016simulation}, weather~\cite{palmer2005representing} and financial market \cite{jeanblanc2009mathematical} modelling, as well as the computer game and gambling industries~\cite{schiffler2010physical}.

Traditionally, random number (RN) generation has been dominated by mathematically complex algorithms that, given a seed, produce a sequence displaying statistical pseudo-random properties \cite{vonNeumann1951}. However, if an attacker had access to the deterministic algorithm as well as the seed, all security would be lost. Therefore, RN generation has undergone a fundamental shift towards focusing on truly unpredictable, non-deterministic processes to extract randomness. Indeed, understanding the source of the unpredictability is of fundamental importance: the initial stochasticism a system displays may in fact stem from an incomplete knowledge of said system or a limitation of the formalism~\cite{Einstein1935EPR,bell2004speakable}. In addition, a flawed (and perhaps malicious), seemingly chaotic but ultimately deterministic RNG, can theoretically output characteristically random strings.

This being so, the inherent non-determinism of quantum physics provides an important source of randomness~\cite{born1955statistical}, which is often exploited to create quantum random number generators (QRNGs). Broadly speaking, two subcategories of QRNGs exist: trusted device QRNGs, in which a simple but fixed system outputs RNs at a generally high rate and low cost. This system, however, assumes that the device's manufacturer is fully trusted. On the other hand, device-independent `self-testing' QRNGs exist, in which the quantum randomness is verified using an entanglement witness or non-local Bell inequality violation. Such devices, however, are often extremely inefficient and require initial random seeds \cite{ma2016quantum}. What is certain, however, is that most QRNGs arise from the field of optics. 
Indeed, in photonics, random bit sequences are often extracted from the Poissonian statistics intrinsic to the photon emission and detection processes~\cite{vincent1970generation,silverman2000tests,ma2005random,stipvcevic2007quantum} (see Ref.~\cite{herrero2017quantum} for a review of RNGs based on these and other optical sources). Furthermore, in the quantum photonics subfield, some non-local (even potentially spacelike-separated) correlations between entangled particle pairs also display intrinsic randomness properties~\cite{masanes2006general, acin2016certified}, such as correlations in polarisation~\cite{fiorentino2007secure}, photon arrival times~\cite{wayne2009photon}, path branching~\cite{jennewein2000fast} and photon number states~\cite{ren2011quantum}.

Although these protocols are effective in generating high quality unpredictable bits, limitations exist. Firstly, optical transmission and detection systems are often inherently asymmetric and introduce unwanted, detrimental bias in the generated bits. This affects the overall randomness of a string. To counter this, post-measurement algorithms do exist to balance the ratio of 0s to 1s (from von Neumann's simple pairwise grouping and discarding protocol~\cite{von195113} to more complex protocols \cite{nisan1996randomness}). However, such unbiasing measures are inefficient and preclude real-time, clean RN generation. So, a system allowing for dynamic control of the bias during the generation process - without compromising the `quantumness' of the protocol - is beneficial (in fact, unbalancing the ratios of 0s to 1s has some important cryptographic applications~\cite{lo2005efficient,xu2015adjustable}). The protocol, presented herein is such a system.

Secondly, some early optical QRNGs extracted only one random bit for each photon measurement event, such as path or polarisation measurements. This, combined with extremely inefficient single photon and entangled photon sources, have deleterious effects on the speed of RN generation which reduce their practicalities. Apart from using more efficient sources and equipment, it would be beneficial to instead consider a degree-of-freedom corresponding to a higher-dimensional Hilbert space. Measuring positions of arrival photons~\cite{yan2014multi} and single qubit quantum walks~\cite{sarkar2019multi} are two such multi-dimensional solutions. The protocol to be outlined below introduces the orbital angular momentum (OAM) degree of freedom as a viable alternative, given its success in other protocols for accessing high-dimensional and multi-dimensional spaces \cite{forbes2019quantum,erhard2020advances}. 

To this end, here we propose a quantum photonic scheme focusing on photons entangled in their OAM degrees of freedom to generate random numbers. We first demonstrate a path branching setup in which, in place of the archetypal beam splitter, a spatial light modulator (SLM) is employed to completely probabilistically direct an incident photon from a down-converted SPDC pair into one of two optical paths (with the other photon functioning as a herald). The use of an SLM allows for projective measurements providing full control by digital holograms over the probability of the incident photon choosing either path. Specifically, one can create and remove bias on-the-fly by scaling the digital hologram masked on the SLM screen.

Next, we consider the potential of using two photon states which are entangled in their OAM degree of freedom. By projecting a photon onto OAM-dependent paths (with the other still acting as the heralding photon), we are able to generate sufficiently random bits given various OAM combinations. The SLM employed in this OAM-based QRNG still easily allows dynamic system unbiasing (based on the initial spiral bandwidth of the down-converted photon pair). Furthermore, OAM opens up the high-dimensional state space. This approach can hence be extended to generate multiple bits per photon detection.

The outline of this paper is as follows: Sec.~\ref{sec:random} gives a brief overview of salient points in studies of entropy and randomness; Sec.~\ref{sec:spatialspdc} outlines the theory behind optical paraxial modes containing orbital angular momentum; after outlining the experimental setup in Sec.~\ref{sec:setup}, we discuss the results in Secs.~\ref{sec:ResultsDiscussion}, finally giving some concluding remarks in Sec.~\ref{sec:conclusion}.

\section{Quantifying randomness}
\label{sec:random}

Before designing and subsequently characterising a quantum random number generator, one needs a basic understanding of the mathematics behind the `randomness' of a string of numbers as well as how to measure its uncertainty. An ideal random string of length $n$, in base $b$, is a sequence of $n$ values independently drawn according to a discrete uniform distribution from the set $[ 0, 1, \cdots, b-1 ]$. The sequence elements necessarily need to be drawn independently of one another: it should be impossible to predict the subsequent element in the sequence, even with sight of all previous elements \cite{NIST}. In base $2$, the situation corresponds with $n$ trials of throwing a two-sided unbiased coin. Despite the fact that humans tend to have an intuitive understanding of the concept of `randomness', its study remains ongoing and there are hence alternative ways of defining and understanding it. For example, \textit{Kolmogorov randomness} posits that a string is only fully random if a computer programme needed to reproduce the string is longer than the string itself \cite{li2008introduction}, whereas information theory typically takes random numbers to be those which maximise a chosen measure of the information entropy of the numbers \cite{goodfellow2016deep}.

A notion closely related to that of entropy in information theoretical contexts is that of \textit{self-information}, which can intuitively be understood as the amount of information learned when observing a value $x$ of a random variable $X$. The pioneer of information theory, Claude Shannon, required self-information to meet a few intuitive axioms \cite{shannon1948mathematical}:
\begin{itemize}
\item An event certain to occur yields no new information: $I(p = 1) = 0$
\item The more unlikely an event is to occur, the more information it's observance gives, with this increase in information being continuous and positive: $I(p_1) \leq I(p_2)$ if $p_2 \leq p_1$, with $I(p) \geq 0 \; \forall \;  p \in [0,1]$
\item The information gained from observing two independent events is the sum of their individual self-informations: $I(p_1 \times p_2) = I(p_1) + I(p_2)$
\end{itemize}

Up to a multiplicative factor, there is a unique function which meets these axioms: given a random variable $X$, with possible outcomes $x_i$ and a probability distribution $P_X\{x_{i}\}=p_{i}$ for $i=1,2,...,d$, the self-information, $I_X(x_i)$, is given by \cite{shannon1948mathematical, goodfellow2016deep}
\begin{equation}
I_X(x_i) = - \log_b(p_{i}).
\end{equation}

For our case, $b = 2$, hence $I_X$ is measured in \textit{bits}. With this in mind, Shannon defined the \textit{Shannon entropy}, $\text{H}(X)$, of the random variable $X$ to be the expectation value of self-information, i.e.
\begin{equation}
\text{H}(X) = - \sum_{i} p_{i}\log_2(p_{i}).
\label{eq:shannonentropy}
\end{equation}
Shannon entropy is a measure of how informative a random variable $X$ is: it can easily be seen that perfectly random events (i.e. those where $p_{i}$ is the same for all outcomes $x_{i}$) maximise H, whereas perfectly predictable events (those for which $p_{i}$ is either always $0$ or $1$) minimise it.



When working in cryptography or randomness extraction, another oft-used entropy measure, is the so-called `min-entropy', defined as
\be
\text{H}_{\text{min}} = - \log_2 (\max_i \{ p_i \}),
\label{eqn:entropymin}
\ee
i.e. the negative base-2 logarithm of the largest probability $p_i$. This yields a worst case bound on the entropy, i.e., the lower limit of randomness that can be extracted. It can be shown that this measure always bounds the Shannon entropy from below and will hence be considered in what follows since it is a more cautious estimate of a system's randomness.

Apart from measures of entropy, statistical hypothesis testing is another widely used method to assess whether a sequence is indeed random~\cite{maurer1992universal,soto1999statistical}. Given a dataset of strings arising from a model of a random number generator, hypothesis testing applies unbiased analytical tests to the dataset to gauge whether the generator itself tends to output sufficiently random strings. We use the term “sufficiently random” since no amount of hypothesis testing can definitively prove that the protocol produces inherently random and irreproducible strings; the tests themselves only offer a simple `pass or fail' evaluation to improve our confidence that the generator is producing numbers that have characteristics reminiscent of randomness~\cite{kenny2005random}.

While many statistical test suites are available, such as John Walker’s ENT~\cite{walker2008ent}, Dieharder~\cite{brown2013dieharder} or the well-known NIST-800 22a Statistical Test Suite~\cite{NIST}, any finite set of tests can overlook hidden correlations. As such, no particular test suite can be deemed complete and no amount of testing on the outputs from a RNG can guarantee its robustness.

Despite this, we employ the NIST Statistical Test Suite (NIST 800-22a) for the pragmatic reason that it is recognised as the industry standard and formulated from comprehensive theoretical and experimental analysis. The measured statistics of each test are converted to a \mbox{\textit{p-value}} using a $\chi^{2}$ reference distribution. These p-values can then be interpreted as the likelihood that an ideal RNG would have generated a sequence less random than the tested sequence~\cite{marton2015interpretation, NIST}. We evaluate the test results by comparing the p-value to a pre-determined significance level $\alpha$, which is set by the required security level of the application at hand. The significance level can be thought of as the probability that the test will deny that a perfectly random sequence is in fact random (i.e. represents the probability of observing a false negative): obtaining a p-value greater than $\alpha$ implies that one accepts the hypothesis that the random number generator produces truly random strings, while a p-value of less than $\alpha$ implies that the random number generator is faulty. We choose a confidence threshold of $\alpha=0.01$, in which case, on average, one in 100 sequences from an ideal RNG will fail the tests by chance. It should be stressed that statistical tests usually only test the outputted sequences (and do not include any modelling of the random number generator itself) and hence, at best, can be viewed as sanity checks against obvious flaws rather than being definitive proof of randomness; it remains essential to assess the generation process behind a random number sequence~\cite{marton2012generation}.

\section{Spatial modes and SPDC}
\label{sec:spatialspdc}


Our proposed random number generator, which operates on the quantum level, measures aspects of pairs of photons entangled via spontaneous parametric down conversion (SPDC) \cite{shih2018introduction}. We choose the well-known Laguerre-Gaussian (LG) modes, $LG_{p}^{\ell}(\bm{q})$, as the complete basis for the vector space of each photon \cite{miatto2011full}. Here, $\ell$ is the azimuthal index (a photon with such an azimuthal index possesses orbital angular momentum of $\hbar \ell$), $p$ the radial index, and $\bm{q} \in \mathbf{R}^2$ is the 2-dimensional wavevector component transverse to the optical ($\hat{z}$) axis. Note that $LG_{p}^{\ell}(\bm{q})$, the momentum space representation of an LG mode, is the Fourier transform of a real space LG mode solution.  With this, the entangled biphoton state $\ket{\psi_{\text{SPDC}}}$ can be written as a superposition of LG modes
\begin{align}
    \ket{\psi_\text{SPDC}} = \sum_{\substack{\ell_s,\ell_i \\ p_s,p_i}} C_{p_s, p_i}^{\ell_s, \ell_i} \ket{\ell_s, p_s} \ket{\ell_i, p_i},
\label{eqn:spdc}
\end{align}
where $\ket{\ell,p} = \int d \bm{q} LG_{p}^{\ell}(\bm{q}) \hat{a}^\dagger(\bm{q})\ket{0}$ is the state of a photon in an LG mode with indices $\ell, p$; $\hat{a}^\dagger$ is the usual creation operator, $\ket{0}$ is the vacuum state, and the subscripts $s, i$ merely distinguish the two photons (called the signal and idler photons for historical reasons). The probability amplitudes $C_{p_s, p_i}^{\ell_s, \ell_i}$ fully characterise the entangled biphoton state. The joint probability of finding the signal (idler) photon in the $\ket{\ell_s, p_s}$ ($\ket{\ell_i, p_i}$) state upon a joint projective measurement, is given by $|C_{p_s,p_i}^{\ell_s,\ell_i}|^2$, where
\begin{align}
    C_{p_s, p_i}^{\ell_s, \ell_i} \propto \int d \bm{q}_s d \bm{q}_i \Phi(\bm{q}_s, \bm{q}_i) \left[ LG^{\ell_s}_{p_s}(\bm{q}_s) \right]^* \left[ LG^{\ell_i}_{p_i} (\bm{q}_i) \right]^*.
\end{align}

The quantity $\Phi$ describes the profile of the initial pump beam profile (usually a Gaussian beam) and the so-called \textit{phase-matching condition} \cite{shih2018introduction}. By conservation of momentum, the OAM of the input pump photon equals the sum of the OAM values of the signal and idler photons it gives rise to, $\ell_p = \ell_s + \ell_i$~\cite{mair2001entanglement}. Furthermore, assuming a Gaussian pump beam (where $\ell_p = 0 = p_p$) and perfect phase matching conditions, the entangled photons have equal but oppositely charged OAM values, $\ell \equiv \ell_s = -\ell_i$.

Finally, our experimental projective measurement process is sensitive only to different OAM charges and hence implicitly projects onto only the $p=0$ radial index. Designating the signal (idler) photon paths with an A (B) subscript, the prepared SPDC state is hence
\begin{align}
    \ket{\psi_\text{SPDC}} = \sum_\ell C_{\ell} \ket{\ell}_A\ket{-\ell}_B,
\end{align}
with $C_{\ell}$ the probability amplitude weighting for the $\ket{\ell}_A\ket{-\ell}_B$ state. Therefore, performing a joint coincidence measurement on the state $\ket{m}_A\ket{-m}_B$ (i.e. projecting onto state $\ket{m}$ and $\ket{-m}$ in the signal and idler arms, respectively) picks out the corresponding amplitude $C_{m}$. Therefore, the probability of measuring the state $\ket{\psi_\text{SPDC}}$ to be in the $\ell^{\text{th}}$ mode is given by $|C_{\ell}|^{2}$.

\section{Experimental setup}
\label{sec:setup}

\begin{figure*}[htb!] 
\centering    
\includegraphics[width=\textwidth]{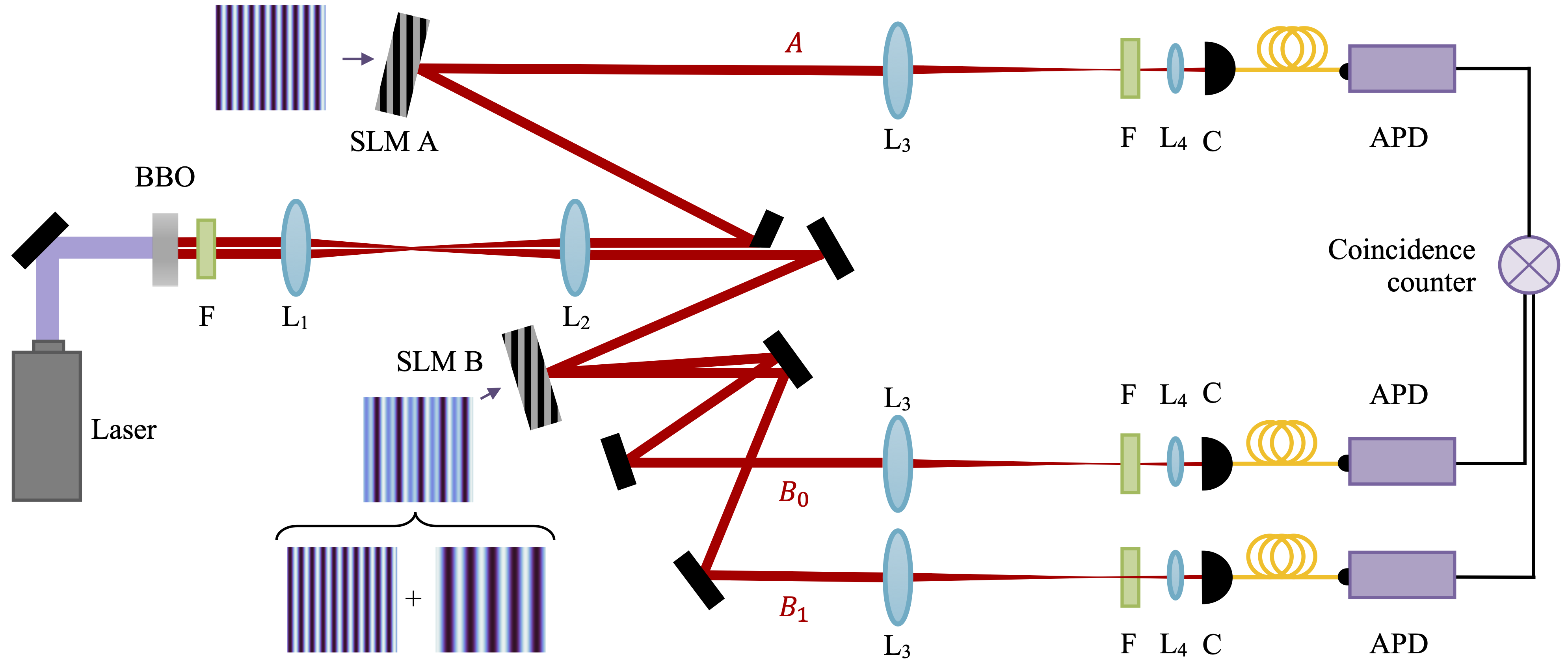}
\caption{After down-conversion, the photon pair is split into two paths. The path A photon acts as the heralding photon for that in path B, which itself is further split into paths $B_0$ and $B_1$, by way of an SLM. A random bit is generated by detecting a coincidence between photon $A$, and either the photon in path $B_0$ or $B_1$. F = filter, $\text{L}_1$ = 200~mm, $\text{L}_2$ = 400~mm, $\text{L}_3$ = 500~mm, $\text{L}_4$ = 2~mm, C = coupler.}
\label{fig:QRNGsetup}
\end{figure*}

Our setup, shown in Fig.~\ref{fig:QRNGsetup}, is a variation of a widely-adopted QRNG scheme based on path branching~\cite{jennewein2000fast}. Originally, this RN generator was built using a beam splitter to create a superposition of two states labelled with path information. To allow for more control over both the incident beam as well as the random number statistics, we instead employed a phase-only Holoeye Pluto-2 spatial light modulator (SLM) to probabilistically split the incident beam into the two paths, mimicking the action of a beam splitter.

A 355 nm mode-locked Vanguard UV laser, directed onto a non-linear $\beta$-barium borate (BBO) crystal, resulted in a pair of degenerate Type-I entangled photons. The entangled photons were then spatially separated into paths A and B using a D-shaped mirror. The crystal plane was imaged onto an SLM placed in each arm.

To separate the modulated light (in the first order) from unmodulated light, a holographic diffraction grating was added to SLM A. Photons in path A functioned as the heralding photons. SLM B, however, was masked with a juxtaposition of TWO diffraction gratings, each with different periodic spacings, instead of a single grating. This causes a photon in path B, incident on SLM B, to be directed down one of two paths, $B_0$ or $B_1$ (see Fig.~\ref{fig:QRNGsetup}). This choice of paths occurs completely probabilistically, with the probabilities themselves controlled by modulating the respective grating depths (i.e. the diffraction efficiencies) of the gratings comprising the juxtaposed SLM B mask: altering the grating depths alters the proportion of incident light diffracted into the first order of either grating.

To be more specific, if one is given a diffraction grating hologram with a maximum possible phase depth of $2 \pi$ within some local region, subsequently scaling the phase range of the pixel at transverse position $(x,y)$ by a factor of $M(x,y)$ (where $M \in [ 0,1 ]$) splits the diffracted light into multiple orders according to~\cite{rosales2017shape}
\begin{equation}
   |c_n(x,y)|^2 = \mathrm{sinc}^2 \left( \pi(n - M(x,y)) \right),
\label{eqn:ampmodulation}
\end{equation}
where $n$ is the diffraction order, and $|c_n|^2$ the fraction of power in said $n^{th}$ order \cite{toninelli2019concepts}. For the first order (\mbox{$n = 1$}), reducing the phase depth across the entire hologram directs a greater proportion of the incident photons out of the first order. Thus, in the case of two juxtaposed gratings corresponding to paths $B_0$ and $B_1$, we can digitally control any bias in the system to achieve the desired ratio of photons in each path. This makes it easy to account for experimental imperfections in the system such as asymmetric detector efficiencies and different optical losses in each arm. Furthermore, this also eliminates the need to consider cumbersome randomness distillation algorithms.

After the SLMs, lenses in the three paths coupled the photons to fibres connected to PerkinElmer avalanche photodiodes (APDs), which output an electronic pulse signal for every detection event observed. A Hydraharp 400 time-tagged each pulse (with a resolution of 1 ps), creating a record of both the time as well as the path in which the detection event occurred. To extract photon coincidences, one then simply reconciles the overlapping APD pulses separately arising from a photon detection in either arm $B_0$ or $B_1$, with a pulse in arm $A$. Photon A, as the trigger, causes either of the other two detectors to register a corresponding entangled photon, if the arrival time lies within a small interval. This is a close approximation to a localised single photon state~\cite{hong1986experimental}.

The resultant data was post-processed to generate random bits: two detectors were said to have detected a pair of entangled photons in coincidence if a single photon was tagged at each respective detector, with the corresponding arrival times differing by a maximum of 25 ns (minimising false coincidence count positives). If detectors $A$ and $B_0$ each registered a photon with the arrival times differing by less than 25 ns, a `0' was appended to our random bit string. Likewise, a `1' was added for a detector $A-B_1$ coincidence. This entire process was fully automated using \textit{LabVIEW}.

Finally, it should be noted that an unbiased system will register the same number of events in each of the $B$ arms, the sum of which should ideally equal the number of events in arm $A$. Also, although the down-conversion efficiency of the crystal and the extraction algorithm itself affect the protocol's efficiency, the most significant bottleneck in the RN generation rate is usually the single photon APDs \cite{ma2016quantum}.

\section{Results and Discussion}
\label{sec:ResultsDiscussion}

\subsection{Random numbers from path information}
\label{subsec:rnpath}

\begin{figure}[htb!] 
\centering    
\includegraphics[width=\linewidth]{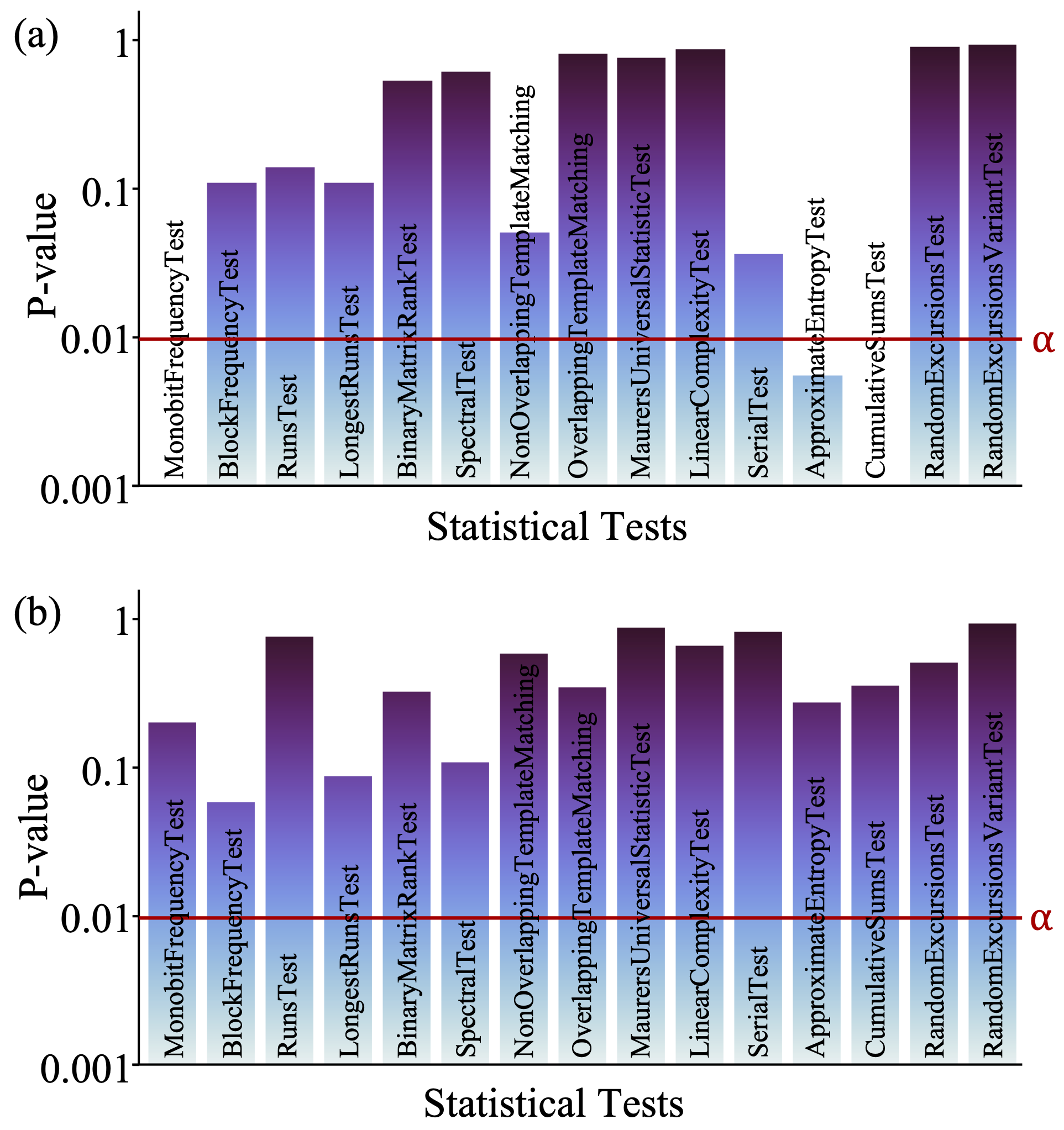}
\caption{ NIST statistical test suite results for a 1 Mb sequence generated by (a) the initial QRNG, with inherent experimental disparities, and (b) the unbiased QRNG, compensated by adjusting the SLM phase grating. P-values above the significance level set at $\alpha=0.01$, pass the test, confirming the absence of patterns in the string.}
\label{fig:NISTresults}
\end{figure}

Random sequences were first generated using the experimental setup of Fig.~\ref{fig:QRNGsetup}. An ideal random sequence should be unbiased in the ratio of its bits. However, this was, perhaps unsurprisingly, not the case initially: experimental disparities between the arms gave unequal coincidence detection probabilities, $p_0$ and $p_1$, between the detector pairs $A-B_0$ and $A-B_1$, respectively. The bias ratio, $R = p_0/p_1$, would be 1 for a perfectly unbiased system. Initially, this ratio was found to be $R = 0.8518 \pm 0.0014$ (so arm $B_1$ was slightly favoured), resulting in a min-entropy of $\text{H}_{\text{min}} = 0.8889 \pm 0.0012$ bits, as per Eq.~\ref{eqn:entropymin} from Sec.~\ref{sec:random}. While a small error in certain contexts, the batteries of randomness tests typically pick up on such small biases of this magnitude. A generated RN sequence (which was 1Mb in length), when subjected to the NIST test suite, failed to pass three of the 15 statistical tests as per Fig. \ref{fig:NISTresults}(a), with p-values well below the chosen significance level of $\alpha = 0.01$ (a level typical for cryptographic applications~\cite{NIST}). Such a discrepancy in the system, even if relatively small, greatly affects the randomness of the generation process, requiring many QRNG protocols to apply numerical unbiasing techniques to extract clean randomness from imperfect data \cite{nisan1996randomness}. Such an extra step is often inefficient and retards the generation rate by necessitating the consumption of some quality random bits.

To overcome this, we alter the diffraction efficiency of the more favoured arm ($B_1$) by changing the phase grating depth of the SLM hologram, as per Eq.~\ref{eqn:ampmodulation}. The SLM employed here was calibrated to display 8-bit grey scale images representative of a total possible phase change of $2\pi$. Scaling the grating depth of arm $B_1$ by a factor $M\in[0,1]$ decreases the probability $p_1$ only, and hence the overall bias. So, $p_1 \to p'_1 = p_1|c_1|^2$, while $p_0 \to p'_0 = p_0$. After renormalising the probabilities such that $p'_0 + p'_1 = 1$ (since scaling $M$ causes photons to leak into higher diffraction orders, which are discarded), we have
\begin{align}
   p'_1 = \frac{p_1 |c_1|^2}{p_0 +p_1 |c_1|^2} =\frac{\text{sinc}^2(\pi(1-M))}{R +\text{sinc}^2(\pi(1-M))}.
\end{align}

\begin{figure*}[htb!] 
\centering    
\includegraphics[width=\textwidth]{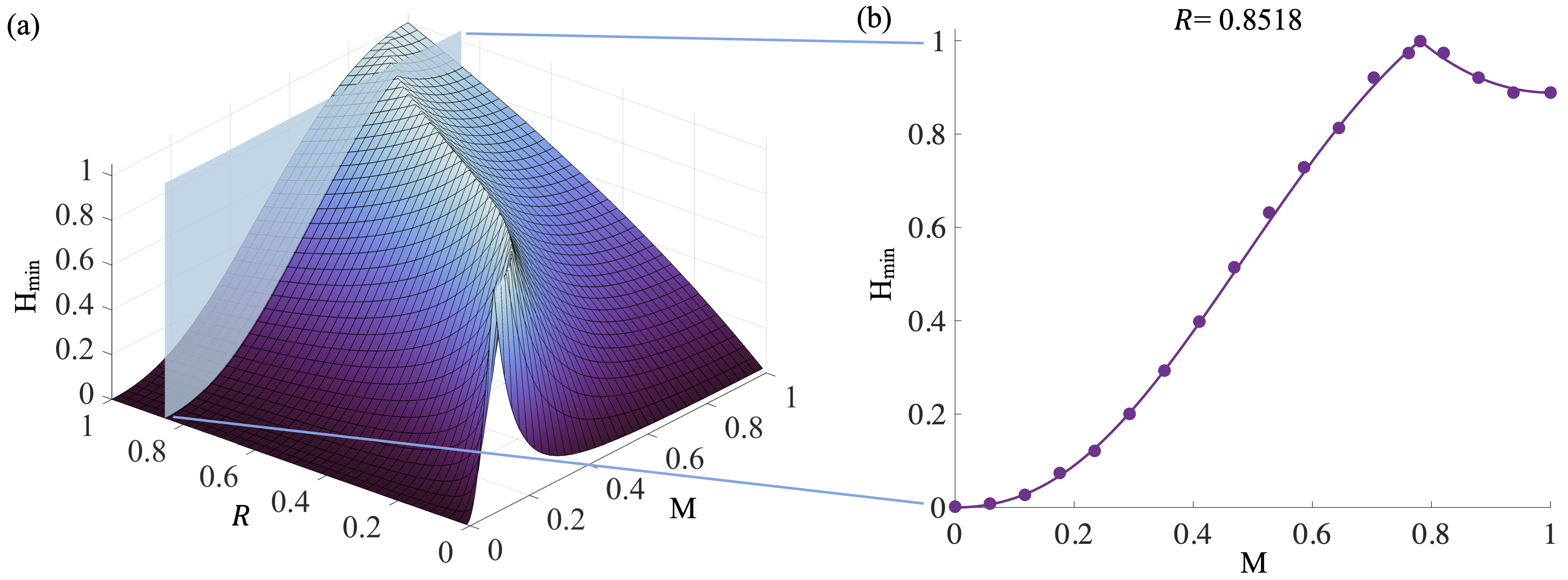}
\caption{(a) The min-entropy theoretically derived from our QRNG as a function of both the grating scaling, $M$, and the bias ratio, $R$. For the initial bias in the system, with $R = 0.8518 \pm 0.0014$, inset (b) shows how the min-entropy depends on the grating scaling. The grating scaling at which the min-entropy is maximised occurs at $M=0.7812$, at which point the min-entropy is $\text{H}_{\text{min}}=0.9991 \pm 0.0003$ bits per bit. This is confirmed by the experimental measurement points; the error bars are too small to be shown at this scale.}
\label{fig:entropyvsgrating}
\end{figure*}

The min-entropy of the altered system is then $\text{H}_{\text{min}} = - \log_2 (\max \{ p'_0, p'_1 \})$. The effect of a change in grating depth, $M$, on the entropy of a system with an initial bias ratio $R$ is depicted in Fig.~\ref{fig:entropyvsgrating}(a), with experimental results shown in Fig.~\ref{fig:entropyvsgrating}(b). The data points in (b) concur with the expected trend of a system with an initial bias of $R = 0.8518 \pm 0.0014$. Given this, to maximise the entropy, the count rate (detection probability) of arm $B_1$ needs to decrease to match that of arm $B_0$. This was calculated to occur when the grating in arm $B_1$ is scaled by a factor of $M=0.7812$. Finally, it follows that since the typical resolution of the SLM is $1/256$ (corresponding to a phase variation of $2 \pi$) for our 8-bit grey scale holograms, we can alter the grating depth of the system to achieve a min-entropy error of approximately
\begin{align}
    \delta \text{H}_{\text{min}} \approx \delta M \left.\frac{\partial \text{H}_{\text{min}}}{\partial M}\right|_{\substack{R=0.8518,\\ M=0.7812}} = \frac{1}{256}\times1.0727 \approx 0.0043.
\end{align}
In practice, the detection probabilities of both arms were altered such that a bias ratio of $R' =p'_0/p'_1 = 0.9988 \pm 0.0014$ was achieved, corresponding to a min-entropy of $\text{H}_{\text{min}}=0.9991 \pm 0.0003$ bits per photon for the corrected system.

The application of this step resulted in a bit bias ratio small enough for the generated string (which, unprocessed, was 1 Mb long) to pass the NIST statistical test suite, Fig.~\ref{fig:NISTresults}(b): the p-values for the various tests were all greater than the chosen significance level of $\alpha = 0.01$. This absence of patterns in the string gives credence to the randomness hypothesis for our random number generator. However, although increasing our confidence in the generator, such statistical tests cannot be used as a comparative measure between different generators.

Our system achieved a bit generation rate of 24 kHz for Gaussian mode detection using single mode fibres, reaching up to 0.46 MHz with multi-mode fibres. Although, optics-based QRNGs have achieved higher bit rates~\cite{ma2016quantum}, such studies often require unbiasing post-processing procedures which significantly reduced their effective bit rate. Our use of SLMs forwent such unbiasing requirements. 
 While the bit rate could be increased using more efficient sources and equipment, it still remains practically limited by the time resolution of the detectors. Furthermore, one of the most efficient ways of increasing the bit rate of a randomness generator is by measuring a high-dimensional Hilbert space~\cite{ma2016quantum}, in which more than one bit of entropy is extracted per detection event. The inclusion of SLMs to dynamically control efficiencies in a higher-dimensional optical QRNG could easily allow for automatic, real-time unbiasing. This in turn potentially allows for the real-time generation of random strings, a feature missing in most contemporary randomness generator studies.

\subsection{Random numbers from OAM}
\label{subsec:rnoam}

Finally, we explore the proposed protocol in which we spatially project onto transverse profiles of single photons to realise a QRNG based on the OAM entanglement of down-converted photon pairs. To do this, the SLM in arm B was masked with superpositions of Laguerre-Gaussian mode holograms and appropriate diffraction gratings to separate the incident photons into OAM-dependent paths (as well as account for inherent bias in the system). By performing these projective measurements on the photons in arm B and coupling only the desired Gaussian mode using single mode fibres, allows us to post-select the desired photon state - a well established spatial mode detection technique~\cite{forbes2016creation}. The detection of the photons in arm A, by way of a multi-mode fibre, ensures that we are able to herald the detection of the photons in arm B without having to modulate photons in arm A and losing the coincidence counts between the detectors.

\begin{figure}[htb!] 
\centering    
\includegraphics[width = \linewidth]{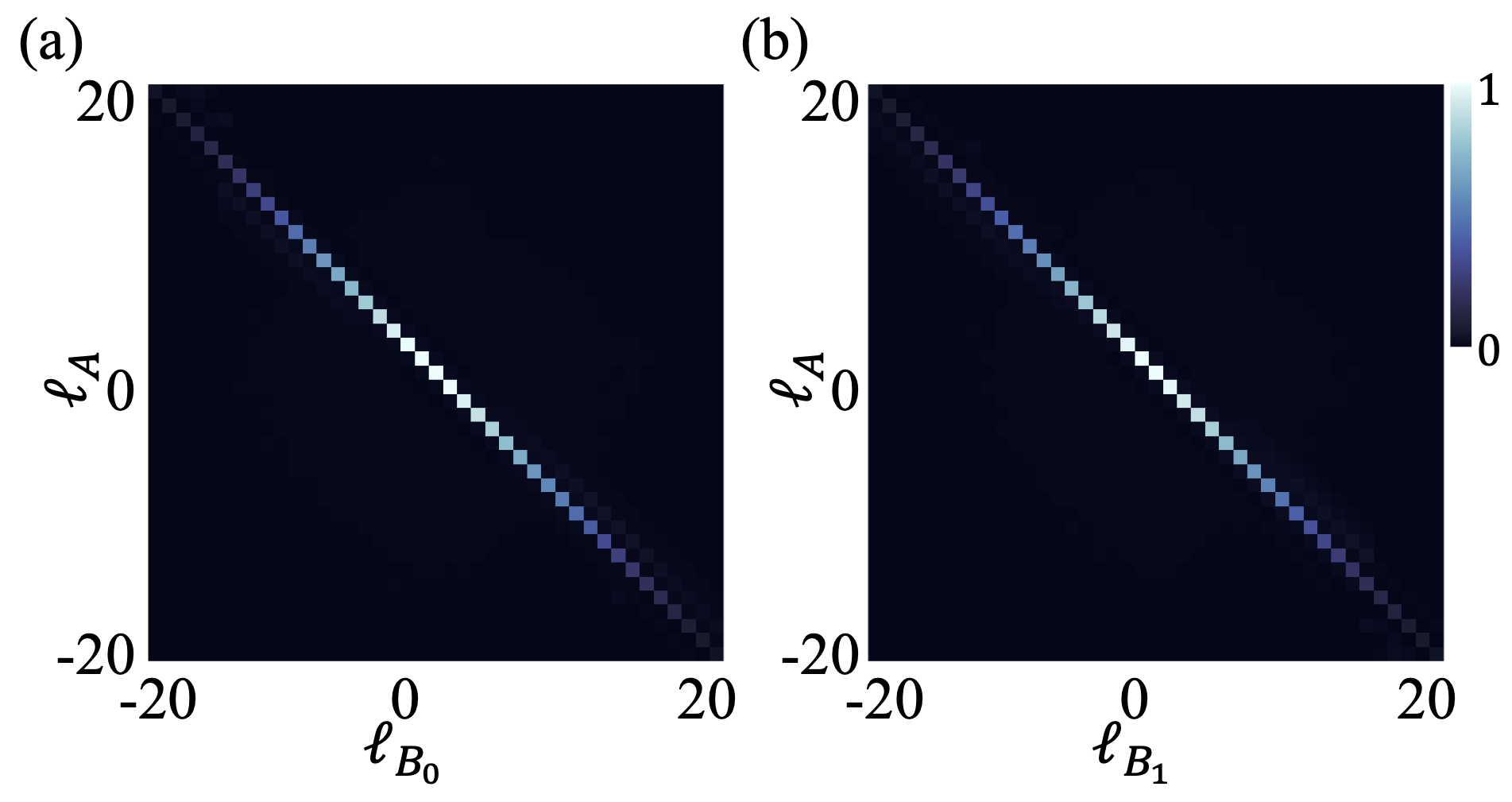}
\caption{Normalised coincidence counts for (a) arm $B_0$ and (b) arm $B_1$, for different projected OAM values. Each arm had a FWHM value of $19$, indicative of a high degree of OAM correlation between the $A$ and $B$ photons.}
\label{fig:QRNGspiralband}
\end{figure}

First, the mode availability in each arm of the RN generator was characterised by spiral bandwidth measurements, i.e., iteratively performing a set of OAM projections in each pair of arms. The normalised coincidence counts for each arm are given in Fig.~\ref{fig:QRNGspiralband}. The full-width-half-max values of $19$ in each arm indicate a large number of the various OAM basis modes were correlated. Using these experimental coincidence counts, it is possible to estimate the entropy of the system given chosen OAM projections in the arms. Indeed, after balancing the juxtaposed gratings on SLM B such that it is equally likely for an incident photon to choose either arm $B_0$ or $B_1$ (as we have shown above), if we project a photon in path $B_0$ onto an $\ell_{B_0}$ eigenstate, and an independent photon in path $B_1$ onto an $\ell_{B_1}$ eigenstate (they're independent since a photon in path B only traverses one of the two subpaths), the min-entropy is $\text{H}_{\text{min}} \equiv \text{H}_{\text{min}}(\ell_{B_0}, \ell_{B_1})$, since the probabilities $p_0, p_1$ in general depend on the modes we project the paths into, namely $\ell_{B_0}, \ell_{B_1}$, in addition to the projection mode in arm A, $\ell_A$. So, $p_0 \equiv p(0|\ell_{B_0}, \ell_{B_1}, \ell_A)$, and similarly for $p_1$. Photon A, however, simply functions as a heralding photon and when estimating the min-entropy conditional on $\ell_{B_0}, \ell_{B_1}$, we in essence use marginal probabilities with the photon A mode `traced out', i.e. summed over, so
\be
p(0/1|\ell_{B_0}, \ell_{B_1}) = \sum_{\ell_A} p(0/1|\ell_{B_0}, \ell_{B_1}, \ell_A)
\ee

\begin{figure*}[htb!] 
\centering    
\includegraphics[width=\textwidth]{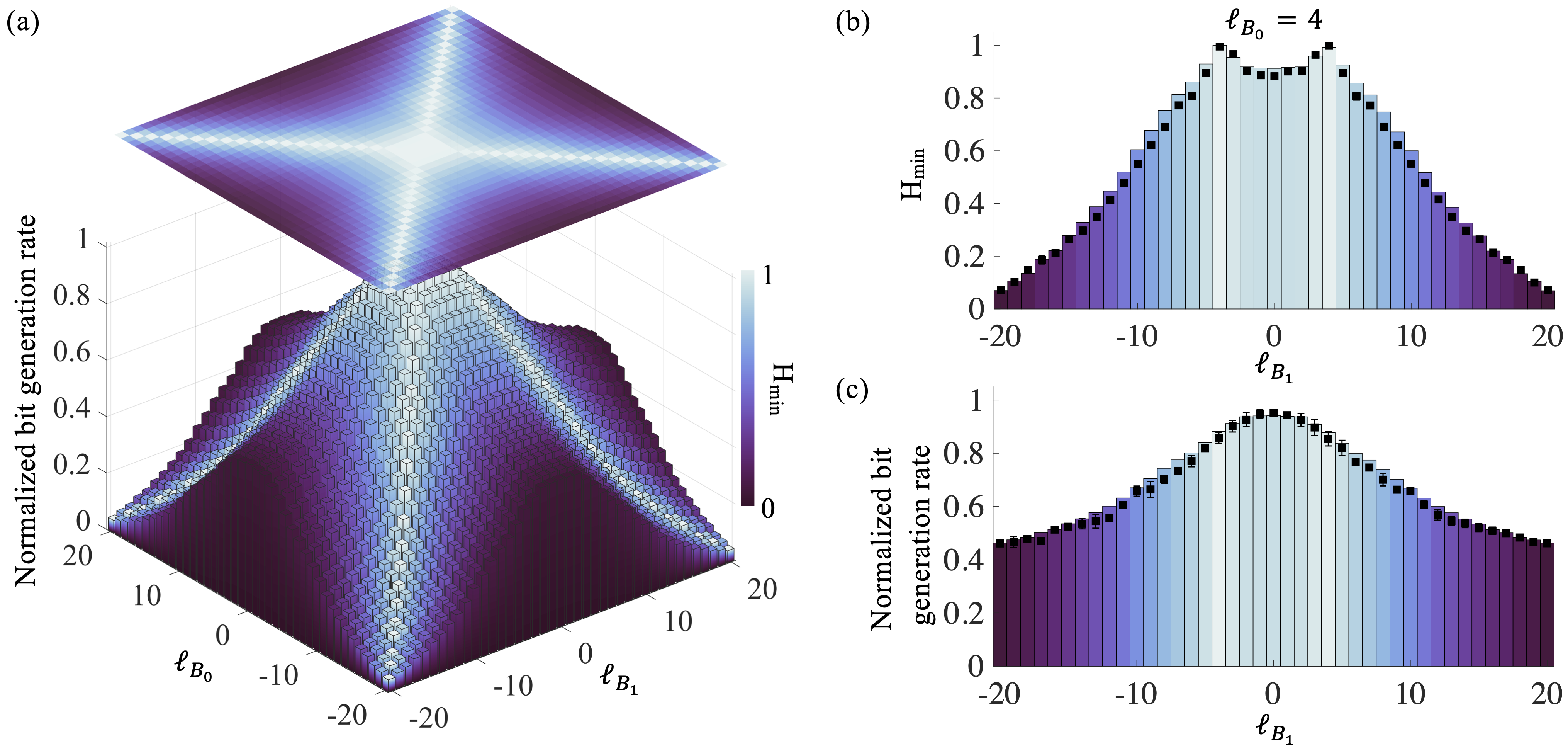}
\caption{(a) Min-entropy (colour) and normalised bit generation rate (z-axis) as function of OAM projection combinations, calculated from the spiral bandwidth measurements. Experimental results for a subset of OAM values, $\ell_{B_0}=4$, $\ell_{B_1}=[-20,20]$, showing the (b) min-entropy and (c) bit generation rates extracted from generated random number strings. The bars correspond to the estimated values from (a). }
\label{fig:entropyvsOAM}
\end{figure*}

Furthermore, the probability of measuring a $0$ does not depend on the chosen projection mode for arm $B_1$, since we only measure a $0$ if the photon traversed arm $B_0$, in which case it doesn't matter what value was chosen for $\ell_{B_1}$. We hence assert that
\be
p(0|\ell_{B_0}, \ell_{B_1}) \equiv p(0|\ell_{B_0}),
\ee
and similarly when observing a $1$. Finally, since the probability of observing a $0$ or $1$ is proportional to the spiral bandwidth coincidence counts of each arm, we can estimate $p(0|\ell_{B_0})$ and $p(1|\ell_{B_1})$ by respectively summing the coincidence counts in Fig. \ref{fig:QRNGspiralband}(a) and (b) over the $\ell_A$ values (assuming a reasonable cutoff) and normalising by their combined sum. Similarly, the estimated normalised bit generation rate is found by summing the coincidence counts over $\ell_A$ for both $\ell_{B_0}$ and $\ell_{B_1}$ contributions and normalising by the maximum bit generation rate of all OAM combinations. Correspondingly, we arrive at the min-entropy values and the generation rates, conditional on $\ell_{B_0}, \ell_{B_1}$, as given in Fig.~\ref{fig:entropyvsOAM}(a).

We experimentally generate various random number strings for a subset of OAM projections, i.e., $\ell_{B_0}=4$ and $\ell_{B_1}=[-20,20]$. The experimental bit generation rates and min-entropy values, are shown in Fig.~\ref{fig:entropyvsOAM}(b) and (c), corresponding to their estimated values. From this, we see that projecting the two B subpaths onto OAM states with equal topological charge magnitudes (i.e. $|\ell_{B_0}| = |\ell_{B_1}|$) results in the highest achievable min-entropy of $1$ bit per detection. This is unsurprising considering OAM momentum conservation, which ensures an equal probability of the conjugate OAM state in either arm. Since the SPDC state roughly follows a normal distribution in the OAM modes, projecting onto higher OAM charges results in a lower detection probability and hence a lower bit generation rate, as in Fig.~\ref{fig:entropyvsOAM}(c). Accordingly, projecting the photon states onto increasingly mismatched OAM values (i.e. states with larger differences in their absolute OAM values) lowers the corresponding bit generation rate and entropy.

Thus, by projecting the photons onto any combination of OAM values one can tailor the ratio of $0$s to $1$s and consequently alter the observed entropy of the system. This comes at the cost of a reduced bit generation rate. One can conceive of a system in which this OAM projection could be used, similar to altering the phase grating depth, to unbias any inherent asymmetries of the system or perhaps to discreetly tailor the ratio of $0$s and $1$s to a desired distribution, in which case a measured sequence NOT following said distribution highlights the potential presence of an adversary. This could be done all the while maintaining the irreproducibility inherent in the quantum system. 

For cryptographic and randomness purposes, often a uniform probability distribution, in which each outcome (here the OAM state) is equally likely to be observed upon measurement. However, the OAM distribution of the down-converted state strongly depends on the profile of the pump field incident on the non-linear crystal. Further advances in the area of pump shaping could be incorporated to finely tailor the distribution and engineer the entropy~\cite{molina2001management, torres2003preparation}.

Finally, given that orbital angular momentum values lie in a higher-dimensional Hilbert space (of dimension $d$), their easy manipulation suggests the possibility of exploiting their higher-dimensional state space to build a multi-bit QRNG. The amount of entropy one could then gain from each photon is $\log_2 d$. Any such demonstration would obviously be contingent on the ability of the system to discriminate between basis modes comprising a $d$-dimensional superposition of OAM states. While a setup similar to the one presented here (namely spatially separating photons with an SLM), a more efficient approach would perhaps entail using mode sorting to conformally map a photon's input OAM charge to a lateral spatial position \cite{berkhout2010efficient} and the subsequent detection of such positions with an array of single photon detectors. This is an experimentally-feasible proposition.

\section{Conclusion}
\label{sec:conclusion}

Our demonstration above was two-fold. Firstly, we demonstrated the integration of digital SLMs for holographic control of the randomness in an otherwise traditional optical QRNG scheme based on path information. This replaced the need for a traditional static beam splitter and easily allowed for the dynamic control of photon path direction and efficiencies. The latter can be easily tailored by simple hologram scaling, offering the ability to remedy any inherent asymmetries in the system and eliminating the need for inefficient or sluggish post experiment randomness extraction techniques. The strings generated from this RNG passed the complete NIST statistical test suite, despite being raw themselves.

Secondly, using holograms composed of gratings and spiral OAM masks, we implemented a QRNG based on the orbital angular momentum degree of freedom of entangled down-converted photon states. The use of spatial light modulators to project onto combinations of OAM charges allowed us to alter the bit generation rate as well as finely tune the bit bias of the generated string, and in effect the entropy of the system while maintaining the irreproducibility of the quantum system.

This work is an important first step in realising an OAM-based, higher-dimensional QRNG. Two properties of an ideal QRNG would be a high bit generation rate, as well as real-time randomness generation (with no randomness extraction procedure required). The protocol here already includes the keys to the former, and modifying it to forgo the projective, iterative `scanning' nature and instead incorporate equipment capable of discriminating the various OAM contribution immediately, opens up the latter.

\section*{Data Availability}
All data is available on request. Correspondence and requests for data should be addressed to AF.

\section*{Acknowledgments}
\label{sec:acknowledgements}
NB acknowledges support from the CSIR DST-IBS programme.

\section*{Author Contributions}
MdO and NB developed the theory. MdO performed the experiments and data analysis; NB wrote the relevant software. AF conceived of the idea and supervised the project. All authors contributed to the writing of the manuscript.


\bibliographystyle{apsrev4-1}
\bibliography{References_File.bib}

\end{document}